\documentstyle[aps,prl,twocolumn,floats,amssymb]{revtex}

\def\spose#1{\hbox to 0pt{#1\hss}}
\def\ltapprox{\mathrel{\spose{\lower 3pt\hbox{$\mathchar"218$}}
 \raise 2.0pt\hbox{$\mathchar"13C$}}}
\def\gtapprox{\mathrel{\spose{\lower 3pt\hbox{$\mathchar"218$}}
 \raise 2.0pt\hbox{$\mathchar"13E$}}}

\begin{document}
\draft

\twocolumn[\hsize\textwidth\columnwidth\hsize\csname@twocolumnfalse\endcsname
\title{
Comment on ``Bicritical and Tetracritical Phenomena
and Scaling Properties of the SO(5) Theory''
}
\author{Pasquale Calabrese$\,^1$, Andrea Pelissetto$\,^2$, 
Ettore Vicari$\,^3$ }
\address{$^1$ Scuola Normale Superiore and  I.N.F.N., Piazza dei Cavalieri 7,
 I-56126 Pisa, Italy.}
\address{$^2$ Dipartimento di Fisica dell'Universit\`a di Roma I
and I.N.F.N., I-00185 Roma, Italy.}
\address{$^3$
Dipartimento di Fisica dell'Universit\`a 
and I.N.F.N., 
Via Buonarroti 2, I-56127 Pisa, Italy.\\
{\bf e-mail: \rm 
{\tt calabres@df.unipi.it},
{\tt Andrea.Pelissetto@roma1.infn.it},
{\tt vicari@df.unipi.it}.
}}

\date{\today}

\maketitle

\begin{abstract}
We show that, in the most general $N$-component theory with symmetry
O($n_1$)$\oplus$O($n_2$), $N=n_1+n_2\geq 3$, the O($N$)-symmetric  fixed point 
has (at least) three unstable directions: the temperature, the
quadratic anisotropy, and the spin-4 quartic perturbation.
This implies that in the SO(5) theory of high-$T_c$ superconductivity
an additional tuning is required to obtain an enlarged 
SO(5) symmetry. This is in contrast with the recent
numerical results reported by  Hu, Phys. Rev. Lett. 87, 057004 (2001).
\vskip 1truecm
\end{abstract}

]


Interesting multicritical phenomena arise from the 
competition of distinct types of ordering \cite{KNF-76}. 
According to the SO(5) theory \cite{Zhang-97}
of high-$T_c$ superconductivity, 
the SO(5) symmetry is realized
at a bicritical point, where two critical lines merge:
one is related to an antiferromagnetic order and 
SO(3) symmetry, the other one to 
a $d$-wave superconducting order and  U(1) symmetry.
Evidence in favor of this scenario has been recently
presented by Hu \cite{Hu-01,Hu-02}, who reported Monte Carlo (MC)
simulations for a five-component spin model. 
He concluded that
the multicritical point describing the
simultaneous SO(3) and U(1) ordering is a bicritical point belonging
to the O(5) universality class
(when the parameters of the model
are in the region relevant for the high-$T_c$ superconductivity). 

Multicritical phenomena in an $N$-component system 
have been studied by considering 
the most general $N$-component Hamiltonian with
symmetry O($n_1$)$\oplus$O($n_2$) containing up to quartic terms,
i.e. \cite{KNF-76}
\begin{eqnarray}
{\cal H}=& \int d^d x \Bigl\{ {1\over 2} \left[
(\partial_\mu \phi_1)^2 
+(\partial_\mu \phi_2)^2+ r_1 \phi_1^2+ r_2 \phi_2^2\right] \nonumber \\
& + u_1 \,\phi_1^4+u_2 \,\phi_2^4+ 2 w\, \phi_1^2 \phi_2^2 \Bigr\},
\label{bicrH}
\end{eqnarray}
where $\phi_1$, $\phi_2$ are $n_1$-,
$n_2$-component fields with $n_1+n_2=N$.
The stability of the fixed points (FP's) depends on the actual values of $n_1,n_2$.
According to the $O(\epsilon)$ analysis of Ref.~\cite{KNF-76},
the bicritical O($N$) FP is stable for 
$N<N_c=4-2\epsilon+ O(\epsilon^2)$;
for higher values of $N$ the stable FP is tetracritical, described either
by the biconal or the decoupled FP.

In the case of the SO(5) theory of superconductors $n_1=3$, $n_2=2$.
The claim of Ref.~\cite{Hu-01} requires the 
stability of the O(5) FP.
This apparently  contradicts the $O(\epsilon)$ analysis of Ref.\cite{KNF-76}.
Moreover, Aharony argued in a Comment \cite{Aharony-02},
using nonperturbative arguments, 
that the decoupled FP is stable. 
Then, assuming the existence of only one stable FP
as suggested by $\epsilon$ expansion,
he interpreted the MC results of Ref.~\cite{Hu-01}
as a  crossover starting
close to the isotropic FP and slowly running
toward the stable FP or away to a first-order transition.
But, as stressed by Hu in his reply \cite{Hu-02}, 
one cannot really exclude the possibility that
two stable FP's may exist with distinct attraction domains,
since arguments based on $\epsilon$ expansion cannot be considered
as conclusive for three-dimensional systems.

This issue can be clarified by an analysis of 
the stability properties of the O($N$) FP under
perturbations.
We consider generic perturbations ${\cal P}_{m\l}$ where $m$ is
the power of the fields
and $\l$ the spin of the representation of the O($N$) group.
For $m=2$ (resp. 4), the only possible values of $\l$ are 
$\l=0,2$ (resp. $\l=0,2,4$). Explicitly,
\begin{eqnarray}
{\cal P}_{2,0}=&& \Phi^2,\qquad  \qquad  
{\cal P}_{2,2}= \phi_1^2-{n_1\over N} \Phi^2 , \\
{\cal P}_{4,0}=&& (\Phi^2)^2,\qquad  \qquad  
{\cal P}_{4,2}= \Phi^2\left[ \phi_1^2-{n_1\over N} \Phi^2 \right],\nonumber \\
{\cal P}_{4,4}=&& \phi_1^2 \phi_2^2- {\Phi^2 (n_1 \phi_2^2+n_2 \phi_1^2)\over N+4}+
{n_1 n_2 (\Phi^2)^2 \over (N+2)(N+4)}, \nonumber
\end{eqnarray}
where $\Phi$ is the $N$-component field $(\phi_1,\phi_2)$.
The perturbations ${\cal P}_{2,0}$ and ${\cal P}_{2,2}$
are always relevant
(with crossover exponents $\phi_{2,0}=1$ and $\phi_{2,2}=1.40(4)$ for $N=5$).
They must be tuned to approach a multicritical point. 
Field-theoretical analyses already applied to systems with
cubic anisotropy \cite{spin4},
based on six-loop and five-loop series respectively in the framework
of the fixed-dimension and $\epsilon$ expansions,
show that the O($N$) FP is also unstable
against the spin-4 perturbation ${\cal P}_{4,4}$ for $N>N_c$ with $N_c\ltapprox 2.9$.
Thus, we conclude that the bicritical O(5) FP is unstable, independently
of the spin-2 perturbation ${\cal P}_{4,2}$. Therefore
the enlargement of the symmetry to O(5) is not realized asymptotically: 
Beside the double tuning of ${\cal P}_{2,0}$ and ${\cal P}_{2,2}$ 
required by a multicritical 
point,
one needs to tune (at least) a further relevant parameter
to recover asymptotically the
 O(5) symmetry.
However, we note that crossover effects due to the spin-4
perturbation ${\cal P}_{4,4}$ are characterized by a small crossover exponent
$\phi_{4,4}=0.145(7)$. This supports 
Aharony's interpretation of the MC results presented by Hu,
in terms of a slow crossover toward either
the stable tetracritical decoupled FP  or
a weak first order transition.  

We finally note that the above-reported arguments also apply to anisotropic 
antiferromagnetic systems. They show that the bicritical O(3)-symmetric  
FP is unstable. In this case the crossover exponent is
even much smaller: $\phi_{4,4}=0.009(4)$.
Moreover, one can argue, using Aharony's argument,
that the decoupled FP is unstable too. 
The stable FP should be the biconal FP,
which is expected to be close to the O(3) FP,
and therefore to have 
critical exponents very close to the Heisenberg ones.

\end{document}